\documentclass[twocolumn,showpacs,preprintnumbers,amsmath,amssymb]{revtex4}
\usepackage{graphicx}% Include figure files
\usepackage{dcolumn}% Align table columns on decimal point
\usepackage{bm, color}% bold math

\begin{document}

\title{Specific heat and upper critical fields in KFe$_{2}$As$_{2}$ single crystals}% Force line breaks with \\

\author{M.\ Abdel-Hafiez}\email{m.mohamed@ifw-dresden.de}
\affiliation{Leibniz-Institute for Solid State and Materials Research, (IFW)-Dresden, D-01171 Dresden, Germany}
\author{S.\ Aswartham}
\affiliation{Leibniz-Institute for Solid State and Materials Research, (IFW)-Dresden, D-01171 Dresden, Germany}
\author{S.\ Wurmehl}
\affiliation{Leibniz-Institute for Solid State and Materials Research, (IFW)-Dresden, D-01171 Dresden, Germany}
\author{V.\ Grinenko }
\affiliation{Leibniz-Institute for Solid State and Materials Research, (IFW)-Dresden, D-01171 Dresden, Germany}
\author{S.-L.~Drechsler}
\affiliation{Leibniz-Institute for Solid State and Materials Research, (IFW)-Dresden, D-01171 Dresden, Germany}
\author{S.\ Johnston}
\affiliation{Leibniz-Institute for Solid State and Materials Research, (IFW)-Dresden, D-01171 Dresden, Germany}
\author{H.\ Rosner}
\affiliation{Max-Planck-Institute for Chemistry of Solid Materials (CPfS), Dresden, Germany}
%\author{L.\ Boeri}
%\affiliation{Max-Planck-Institute for Solid State Research,  D-70569 Stuttgart, Germany}
\author{A.U.B.\ Wolter-Giraud}
\affiliation{Leibniz-Institute for Solid State and Materials Research, (IFW)-Dresden, D-01171 Dresden, Germany}
\author{B.\ B\"uchner}\affiliation{Leibniz-Institute for Solid State and Materials Research, (IFW)-Dresden, D-01171 Dresden, Germany}
\date{\today}

\begin{abstract}
We report low-temperature specific heat measurements for high-quality
single crystalline KFe$_{2}$As$_{2}$
%single crystals 
%with a sharp superconducting transition at
($T_c \approx $ 3.5\,K). The investigated zero-field specific heat data 
yields
an unusually large nominal Sommerfeld coefficient 
$\gamma_{\rm n}=94(3)$~mJ/mol K$^2$ which is
however 
%might be 
%significantly 
affected by extrinsic contributions as 
evidenced by a sizable residual linear specific heat
and various theoretical considerations
including also an analysis of Kadowaki-Woods relations.
Then KFe$_{2}$As$_{2}$ should be classified as a weak
to intermediately strong
coupling
superconductor with a total electron-boson coupling constant
$\lambda_{\rm tot} \sim 1$ (including a calculated weak electron-phonon coupling constant of
$\lambda_{\rm el-ph}=0.17).$
 From
%detailed 
specific heat and ac susceptibility studies in
external magnetic fields the magnetic phase
diagram has been constructed. 
We confirm the 
high anisotropy of the
%The 
upper critical
fields $\mu_0H_{\rm c2}(T)$
ranging from a factor of 5 near $T_c$
to a slightly reduced value approaching $T=0$
for fields
%have been extrapolated} 
%using a single-band
%Werthamer-Helfand-Hohenberg (WHH) model 
%for $H \parallel c$ and 
$B \parallel ab$ and $\parallel c$
and show that their ratio $\Gamma$ slightly exceeds
the mass anisotropy of 4.35 derived from our
full-relativistic
LDA-band structure calculations. Its slight reduction when
approaching $T=0$ is not a consequence of Pauli-limiting as
in less perfect samples but point likely to a multiband effect.
We also report 
%the 
irreversibility field data
obtained from ac susceptibility
measurements. The double-maximum in the $T$-dependence of its
%the 
imaginary part 
%of the AC susceptibility 
for fields $B \parallel c$ indicates a
peak-effect
in the $T$-dependence of 
%a 
critical currents.
\end{abstract}

\maketitle

\section{Introduction}
Since the discovery of superconductivity (SC) in 
an electron doped LaFeAsO (La-1111)
compound with a superconducting transition temperature $T_{c}$
$\sim$ 26\,K \cite{Kamihara2008}, iron pnictides are of great
interest in fundamental 
%and applied research in 
condensed matter
physics due to their large variety of structural, magnetic and
electronic properties. 
In order to understand the nature of superconductivity in
Fe pnictides a huge amount of 
%different 
theoretical and experimental studies has
been performed but nevertheless
%~\cite{Mazin22009,Ishida2009}. However, still
many 
%open 
questions have not been  answered yet, such as the symmetry of
the order parameter and the pairing mechanism 
%in these compounds 
as
well as their relation to the magnetic properties. In this situation,
low-$T$ specific heat and magnetic susceptibility
measurements might be helpful since
they provide insight into many-body physics via the renormalization of
%provide 
such physical quantities as: the Sommerfeld coefficient
$\gamma_{\rm el}$ (a measure of the renormalized density of states),
the irreversibility field $H_{irr}$, the upper critical field $H_{c2}$, its anisotropy
etc.\ all are important factors 
which  affect the
%o address the nature of
superconducting and the normal state properties as well.
In particular,
%nd the pairing mechanism. 
%Moreover,
 they can shed light on
%.g. 
the Fermi surface topology and other 
relevant
aspects of the electronic
structure. 
To address  the role of magnetism
in the formation of the superconducting state 
%in Fe-basedsuperconductors, 
the heavily hole doped KFe$_{2}$As$_{2}$ is
%nteresting 
worth
to be studied
% This is 
due to its distinctive
characteristics with respect to other 
stoichiometric 122 and 1111
Fe-pnictide compounds:
(i) there is no
%oindications of any 
static magnetic ordering in the sense of an ordinary spin-density wave (SDW) or an orthorhombic 
structural transition
%were observed in measurements of various physical properties of
%oly- and single-crystalline KFe$ _{2}$As$_{2}$ samples
\cite{Fukazawa2009,Kihou2010}. (ii) Superconductivity occurs
in relatively dirty samples near
%the stoichiometric compound with a T$_{c}$ around 
2.8 K
\cite{TERASHIMA2009} 
and increases in cleaner 
high-quality single crystals
up 3.5 to 3.7~K \cite{Kihou2010}. 
% (iii) High quality single crystals of
%Fe$_{2}$As$_{2}$ can be grown . 
(iii)
 Remarkably, no
nesting of the Fermi surface was detected in contrast 
to e.g.\ Ba$_{0.6}$K$_{0.4}$Fe$_{2}$As$_{2}$ \cite{Sato2009,Evtushinsky2011}. However, a
neutron scattering study of heavily hole doped superconducting
KFe$_{2}$As$_{2}$ revealed well-defined low-energy incommensurate
spin fluctuations at $[\pi(1 \pm 2 \delta), 0]$ with $\delta$ =
$0.16$ \cite{Lee2011}. This is different from the previously
observed commensurate antiferromagnetism (AFM) of electron-doped
AFe$_{2}$As$_{2}$ (A = Ba, Ca, or Sr) at low energies. Additionally,
de Haas-van Alphen \cite{Terashima2010} and cyclotron
resonance \cite{Kimata11} studies of
KFe$_{2}$As$_{2}$ revealed a strong mass enhancement of
the quasi-particles. 
%Overall, although KFe$_2$As$_2$ belongs to the
%122 family, its physical properties resemble more those of the 111
%family like LiFeAs \cite{Morozov2010}. 
Furthermore, it exhibits a
very
large anisotropy as compared with less hole-doped 
members of the 122-family and other Fe-pnictide and chalgogenide
supperconductors
~\cite{U2011} After naturally electronically more anisotropic 1111 
and Tl$_{1-y}$Rb$_{y}$Fe$_{1-\delta}$Se$_{2}$ with $y \sim 0.4, 
\delta \sim 0.3$  
superconductors showing only slightly larger or comparable
slope anisotropies of $\sim $ 5 to 6 K-122 belongs to the most anisotropic
pnictides \cite{Zhang2012}.
%heir higher absolute $H_{c2}(0)$-values are plausible
%n view of their much higher $T_c$-values but 
A complete understanding
of their critical field
slopes near $T_c$ is still missing due to the complex interplay
of pair-breaking impurities and the symmetry of the 
superconducting order parameter \cite{Kogan2009}.
%with a very low upper critical field
%compared to much higher values of other Fe-based superconductors.

The magnetic phase diagram of KFe$_2$As$_2$ has been studied
via resistivity
measurements on single crystals \cite{TERASHIMA2009}, however, its
determination using thermodynamic bulk techniques on single
crystalline material is lacking up to now \cite{note1,Kim2011}. 
In the context of recently discovered very high
surface upper critical field $H_{c3}$ with $H_{c3}(T)/H_{c2}(T) \sim 4.4$
for the external field $\parallel$ to the $ab$-plane
in K$_{0.73}$Fe$_{1.68}$Se$_{2}$ \cite{Tsindlekht2011} is noteworthy,
because for that direction the nucleation starts already 
much higher 
at $H_{c3}$
and resistivity and or/ac susceptibility measurements might in principle
lead to a confusion between $H_{c2}$ and $H_{c3}$ suggesting this way an 
overestimated anisotropy of the upper critical fields. Hence, for this 
geometry,   
%In this
%situation, 
specific heat studies of high-quality single
crystals are mandatory to address their bulk
anisotropy.
% as well as details
%of the gap symmetry in an accurate way.

%In this work 
Here we present low-$T$ specific heat and
ac magnetization studies on high-quality
%heavily hole-doped 
superconducting KFe$_{2}$As$_{2}$ single crystals
with a larger $T_c$ and a much higher residual 
resistivity ratio as compared to the first single crystals used for an 
upper critical field study
for KFe$_2$As$_2$ by Terashima {\it et al.} \cite{TERASHIMA2009},
where a large anisotropy ratio of the upper critical fields as well as for the
electric resistivity perpendicular
and parallel to the $ab$-plane have been reported.
The obtained data are analyzed within the framework of various
theoretical approaches. In particular, we found
good agreememt with the electron mass anisotropy
derived from density funtional theory (DFT) considering in- an out-of
plane
plasma frequencies and Fermi velocities. 
%The superconducting transition

\section{Experimental}

Single crystals of KFe$_{2}$As$_{2}$ have been grown using a
self-flux method with K:Fe:As in the molar ratio of 1:4:4. All
preparation steps like weighing, mixing, grinding and storage were
carried out in an Ar-filled glove-box. As a first step the
appropriate amounts of the
precursor materials FeAs and Fe$_{2}$As were
thoroughly grinded in an agate mortar, secondly the exact amount
of weighed metallic K was deposited at the bottom of an alumina
crucible, where on top of it the well grinded mixture is placed
carefully, finally sealed in a niobium crucible. The sealed
crucible assembly is placed in a vertical furnace, heated up to
1373\,K and cooled down to 1023\,K with a rate of 2 K/hour.
Finally the furnace is cooled very fast from 1023\,K to room
temperature. All crystals are
grown with layer-like morphology and they are quite easy to cleave
along the $ab$ plane. The quality of the grown single crystals was
assessed by complementary techniques. Several samples were
examined with a Scanning Electron Microscope (SEM Philips XL 30)
equipped with an electron microprobe analyzer for a
semi-quantitative elemental analysis using the energy dispersive
x-ray (EDX) mode. The composition was estimated by averaging over
several different points of the platelet-like single crystals and
is found to be consistent and homogeneous with a 122 structure
within the instrumental error bars. Typical crystal sizes with a
rectangular shape were about 1.2$\times$0.5 mm$^2$ and a thickness
of 50\,${\mu}$m along the {\it c}-axis. All crystals 
exhibit very similar $T_c$-values (see the insets of Figs.\ 2 and 8)
within the experimental error of about 0.25 K.

Low-temperature specific heat and ac magnetization 
have been
%was
determined using a Physical
Property Measurement System (PPMS from Quantum Design). The specific
heat data were measured using a relaxation technique. For the
measurements at $H \parallel ab$ a small copper block has been used to
mount the sample on the specific heat puck. The heat capacity of the
copper block was determined in a separate measurement and its value has been 
subtracted from the raw data of KFe$_2$As$_2$. 

\section{Results and discussions}
\subsection{ac magnetization measurements}

\begin{figure}
\includegraphics[width=20pc,clip]{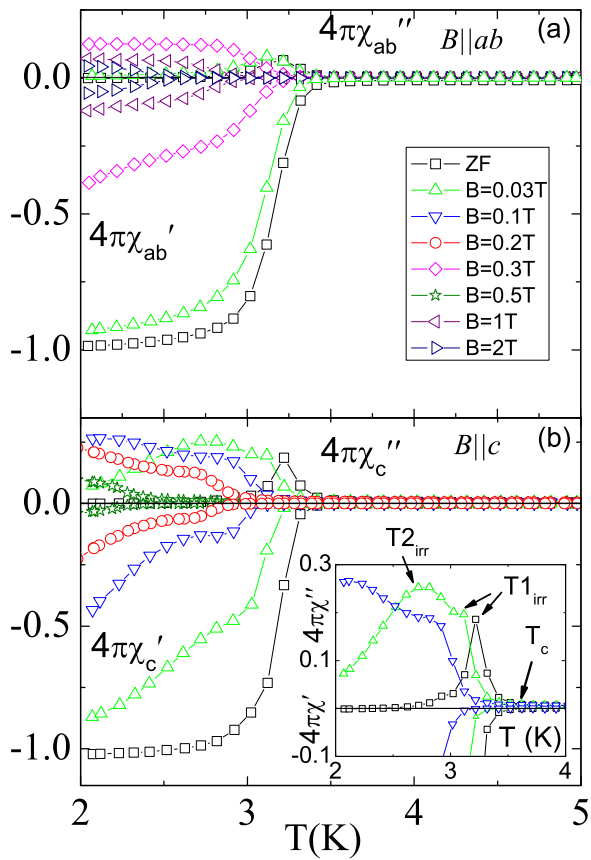}
\caption{(Color online) (a)
The $T$-dependence of the complex ac susceptibility
components
$4\pi\chi{'}_{v}$ and $4\pi\chi{''}_{v}$ of
KFe$_{2}$As$_{2}$ has been
measured in an ac field with an amplitude 5 Oe and a frequency of 1 kHz upon
warming in different DC magnetic fields after cooling in zero
magnetic field with (a)  
$B\parallel ab$ and (b) $B\parallel c$. The sharp
superconducting transition with $\sim$ 100~\%
superconducting volume fraction indicates the bulk nature of
superconductivity and the high quality of our crystal. The inset shows the
criteria used to obtain $T_c$ and $T_{irr}$; for details see text.}
\label{Fig:1}
\end{figure}

Figure~\ref{Fig:1} depicts the $T$-dependence of
the volume ac susceptibilities ($\chi{'}_{v}$ and $\chi{''}_{v}$) of our
KFe$_{2}$As$_{2}$ single crystal. The measurements were done
in an ac field with an amplitude $B_{ac}$ = 5 Oe, a frequency $f$ = 1 kHz and DC fields up to 5 T
parallel to the $ab$-plane as shown in Fig.~\ref{Fig:1}a and parallel to the
$c$ axis in Fig.~\ref{Fig:1}b. Special care has been taken to correct the
magnetization data for demagnetization effects, where the
demagnetization factor has been estimated based on the crystal
dimensions \cite{Osborn1945}. The ac susceptibility measurements can be used for
an investigation of the flux
dynamics in superconductors \cite{Giapintzakis1994,Gomory1997, Vlakhov1994}.
The imaginary part $\chi{''}_{v}$ is related with
the energy dissipation in a sample and the real part $\chi{'}_{v}$ is related
with the amount of screening. Both these
functions depend on the ratio between skin depth $\delta_s$ and the sample
dimension $L$ in the direction of the flux penetration. In the normal state
$\delta_s\sim(\rho_n/f)^{0.5}$, where $\rho_n$ is the normal state resistivity and \emph{f} is the frequency.
In the superconducting state the skin depth $\delta_s\propto\lambda_L$ if an
external magnetic field is below the first critical field $H_{c1}$, where
$\lambda_L$ is the London penetration depth. For magnetic fields above $H_{c1}$,
$\delta_s\propto L_B$, where $L_B\sim\ B_{ac}/J_c$ is the Bean's penetration
depth and $J_{c}$ is the critical current density.
In general, if $L\ll\delta_s$ an ac field completely penetrates the sample
and thus the susceptibility is small. In the case, if $L\gg\delta_s$,
 most of
the sample volume is screened, therefore, $4\pi\chi{'}_{v}=-1$ and
$\chi$$''$$_{v}$ $\rightarrow$ 0.
In accordance with this the ac susceptibility data measured at low $T$ confirm 
the 
bulk superconductivity of our
KFe$_{2}$As$_{2}$ single crystal (Fig.~\ref{Fig:1}). 
%The transition temperature
$T_{c}$ $\sim$ 3.6(1)K has been extracted from the bifurcation point between
$\chi{'}_{v}$ and $\chi{''}_{v}$ susceptibilities as shown in the inset 
of Fig.\ \ref{Fig:1}b. This point is related with a change in the linear resistivity
due to the superconducting transition. It can be also used for the determination
of the $T$-dependence of the upper critical field $H_{c2}$ from ac
susceptibility data measured at various DC fields \cite{Gomory1997}.

At T $<$ $T_c$ the function $\chi{''}_{v}$
increases with decreasing temperature and some value of T = $T_{irr}$ where ($L\sim\delta_s$) $\chi{''}_{v}$
has a maximum (see Fig.~\ref{Fig:1}). The above discussed Bean's approximation
predicts that the $T$-dependence of the peak in $\chi{''}_{v}$ follows
the $T$-dependence of the critical current density $J_{c}$. Thus,
one might
%can 
relate $T_{irr}$ with an irreversibility temperature
and use its DC field dependence to obtain 
%an 
the 
irreversibility field $H_{irr}$
\cite{Vlakhov1994}. 
However, we 
%We 
remind the reader that 
%a such defined
$H_{irr}$ defined this way
is not the 'true' irreversibility field since by definition $H_{irr}$
is the field
at which $J_{c}$ = 0. From this point of view 
it is better to
%one can 
use dc magnetization
data to obtain the irreversibility line.  {In general, we
observed a rough agreement between $H_{irr}$ obtained from dc and ac
susceptibility data but the large value and the strong $T$-dependence of the
normal state dc susceptibility lead to a large uncertainty in determination of
the $H_{irr}$. Therefore, to obtain the irreversibility line we used ac
susceptibility data. (The DC susceptibility data will be presented elsewhere
\cite{Grinenko2012}). Thus}, with some cautions we relate the maximum in the
T-dependence of $\chi{''}_{v}$ to $T_{irr}$. It can be seen
in Fig.~\ref{Fig:1}b that at non-zero $B_{DC}||c$ the single maximum
in  $\chi{''}_{v}$ splits into two features at $T1_{irr}$ and $T2_{irr}$.
Therefore, for $B_{DC}||c$ we defined two
different 'irreversibility' fields  $H1_{irr}$ and  $H2_{irr}$.
The T-dependence of these fields is plotted in Fig.\ 
~\ref{Fig:5}.
\begin{figure}[b]
\includegraphics[width=18pc,clip]{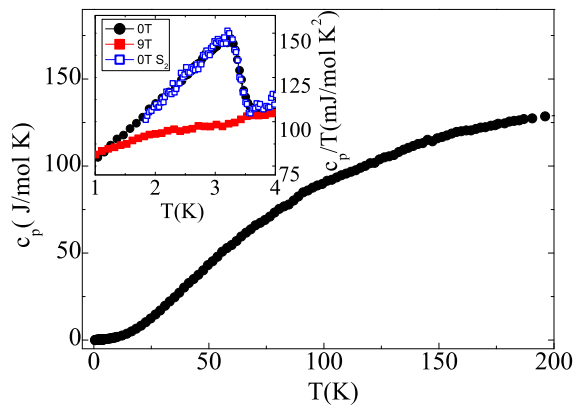}
\caption{(Color online) The $T$-dependence of the zero-field specific heat of
KFe$_2$As$_2$ for 400\,mK $\leq$ $T$ $\leq$ 200\,K. The inset shows
$c_{p}/T$ versus $T$ of our 0\,T and 9\,T data together with the zero-field of 
another sample (S$_{2}$) shows the same $T_{c}(H)$ behavior 
down to 1.8K.}
\label{Fig:2}
\end{figure}

\subsection{Specific heat studies}
Figure 2 shows the $T$-dependence of the zero-field
specific heat measured down to 0.4\,K. A clear 
%and 
sharp anomaly was
observed near 
%around $\sim$ 
3.5 K in agreement with the magnetization
data. In order to determine the zero-field normal state Sommerfeld
coefficient $\gamma_{\rm n}$, the specific heat can be plotted for $T >
T_c$ as $c_p/T$ versus $T^2$ following
\begin{equation}
c_{\rm p} = \gamma_{\rm n}T + \beta_3 T^{3} + \beta_5 T^5 + ... \ ,
\end{equation} 
with $\gamma_{\rm n}$ and $\beta_3$, $\beta_5$ as the nominal electronic
and lattice coefficients, respectively. The obtained values for our
KFe$_2$As$_2$ sample are $\gamma_{n}$ = 94(3) mJ/mol K$^2$ and
$\beta_3 = 0.79$~mJ/mol$\cdot$K$^4$ and $\beta_5=6.09\cdot 10^{-4}$mJ/mol K$^6$. 
Our $\gamma_{\rm n}$ value 
%is
compares very well with
%comparable to 
$\gamma_n$ = 93 mJ/mol K$^2$ reported in 
%of a single crystal sample KFe$_{2}$As$_{2}$ from
Ref.~\cite{Fukazawa2011}. From the relation for
%of 
the Debye temperature, \emph{$\theta_{D}$ = (12$\pi$$^{4}$RZ
/5$\beta_3$) $^{1/3}$}, where \emph{$R$} is the molar gas constant and
\emph{$Z$} = 5 is the number of atoms per formula unit, we obtain
\emph{$\theta_{D}$} = 214~K. 

Notice the enhanced
value of
%Clearly, a strongly
%enhanced value 
$\gamma_{\rm n}$ 
%is present 
%for our heavily hole doped
%compound 
KFe$_2$As$_2$ as compared to other stoichiometric 
and nonstoichiometric
122 compounds  or to any other
superconducting iron pnictide or chalcogenide to the best of our 
knowledge.
The low values of $\gamma_{\rm n}$ for 
BaFe$_{2}$As$_{2}$ and
SrFe$_{2}$As$_{2}$ are not surprising since large parts of its
Fermi 
surface are gapped due to the well-known magnetic spin density wave
(SDW)  transition at high
temperatures. Hence, in these cases a comparison with a hole doped
system where the SDW transitions are suppressed is more meaningful.
For instance, for the closely related, nearly optimal 
hole-doped systems 
Ba$_{0.68}$K$_{0.32}$Fe$_{2}$As$_{2}$  ($T_c= 38.5$~K) \cite{Popovich2010}, 
Ba$_{0.6}$K$_{0.4}$Fe$_{2}$As$_{2}$  ($T_c= 36.5$~K) \cite{Mu2009} or 
Ba$_{0.65}$K$_{0.35}$Fe$_{2}$As$_{2}$  ($T_c= 29.4$~K) \cite{Pramanik2011},
the Sommerfeld parameters
$\gamma_{n}=50.0$, 63.3 and 57.5~mJ mol$^{-1}$K$^{-2}$, respectively,
have been reported. 
\begin{figure}[b]
\includegraphics[width=16pc,clip]{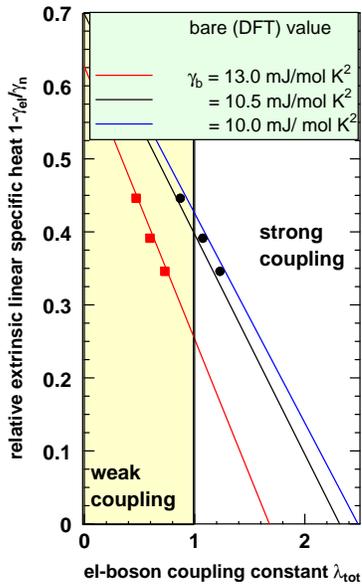}
\caption{(Color online) The relative extrinsic 
linear specific heat coefficient of KFe$_2$As$_2$ 
using the experimentally observed nominal
$\gamma_{\rm n}=94$~mJ/mol $\cdot$K$^2$
vs.\ 
the
total electron-boson 
coupling constant $\lambda_{\rm tot}$ given by Eqs.\ (4,7)
for various bare $\gamma_{\rm b}$-values obtained from 
density of states as calculated by various
DFT codes [22,25,26] 
%\cite{Hashimoto2010,Rosner,Popovich2010} 
(see text) and using 
a typical high-energy renormalization factor of $\eta=2.7$ 
(see Eq.\ (4)). The data points show the results of simulations within 
single band $d$-wave Eliashberg theory to reproduce $T_c=3.5$~K and a spectral
density for spin fluctuations
adopted from recent INS data [7] 
%\cite{Lee2011} 
and including also a weak
electron-phonon interaction and a weak Coulomb pseudopotential $\mu^*$
(details will be discussed elsewhere).
}.
\label{Fig3a}
\end{figure}
In view of their much higher $T_c$-values, 
ascribed by the authors  
to an essential contribution from 
strong coupling corrections with $\lambda \sim 2 $ \cite{Popovich2010}
for the 
electron-boson coupling constant
(spin fluctuation mediated interband coupling)
and a comparable bare value of 
$\gamma_{\rm b}\sim$ 10 mol$^{-1}$K$^{-2}$ (see below)
 according to DFT-band structure calculations  
the unusual large value for KFe$_2$As$_2$ reported above provides a 
surprising
puzzle. However, it can be somewhat reduced, if there is an essential
{\it extrinsic} contribution to the 
system of intinerant charge charriers, e.g.\ due to  defect states 
with low-energy excitations 
\cite{Grinenko2012}. To be consistent with that analysis
%Anyhow, to get a more realistic value, we will
we are forced to
assume, that KFe$_2$As$_2$ is 
{\it not} in the strong coupling limit which
seems to be natural in view of its
low $T_c$-value 
(to be discussed within the framework of Eliashberg-theory elsewhere).
The general situation, independent on the strength of the
electron-boson 
coupling regime
and the symmetry of the order parameter, is plotted in Figs.\ 3 and 4
using the calculated 
DFT-values $\gamma_b$ \cite{Hashimoto2010,Rosner,Popovich2010}
as convenient bare values. Thereby the high-energy renormalization
$\eta=2.7$ as derived from the calculated DFT plasma frequencies
of 2.58~eV \cite{Hashimoto2010} and 2.56~eV \cite{Rosner,drechsler09} and
and the expected experimental
unscreened plasma frequency of about 1.55~eV have been taken into account
\cite{remarkopt}.
\begin{figure}[t]
\includegraphics[width=18pc,clip]{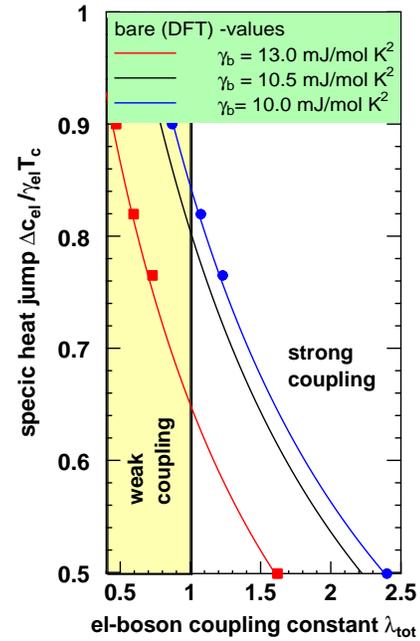}
\caption{(Color online) The normalized specific heat jump at $T_c$
vs.\ the phenomenological total electron-boson coupling
constant $\lambda$ and various DFT-derived bare linear specific heat
values [22,25,26] 
%\onlinecite{Hashimoto2010,Rosner,Popovich2010}. 
%In the analysis a high-energy mass renormalization
%by a factor of 2.7
%determined  from the calculated and measured unscreened plasma frequencies
%of 2.62~eV and 1.6~eV, respectively,  has been taken into account.%
Red and blue symbols: results using the 
corresponding
$\gamma_{\rm el}$ values from Fig.\ 3 extended by two strong coupling
points for $\gamma_{\rm el}=\gamma_{\rm n}$
and the experimental value of 
$\Delta c_{\rm el}/T_c=45.6$~mJ/ mole $\cdot$K$^3$)}.
\label{Fig:4}
\end{figure}
There are at least two experimantal hints which clearly point to the existence
of such an extrinsic subsystem which manifest itself in 
a substantial residual linear specific heat visible at very low-$T$
at ambient fields (see Fig.\ 8) and in high fields 
of about 9~T where the superconductivity is well suppressed
(see inset of Fig.\ 2). From the latter one estimates 
$\gamma_{\rm el} \leq 70$~mJ/mol K$^2$.
In the next section we will provide theoretical arguments in favor
of a significantly reduced intrinsic Sommerfeld term $\gamma_{\rm el}$.

\subsection{Theoretical estimates of the thermal mass enhancement}
\subsubsection{Kadowaki-Woods analysis}

The weak
coupling
 result can be understood at a qualitative level also by analyzing
the so called Kadowaki-Woods relation (KWR)
$\kappa_{\rm KWR}=A_{\rho}/\gamma_v^2$, where 
$A_{\rho}$ describes the $T^2$-contribution to the resistivity
at very low $T$:
$\rho(T)=\rho_0 +A_{\rho}T^2$
observed so far only in few very clean samples \cite{Hashimoto2010}
with  extremely large residual resitivity ratio 
$\rho(300K)/\rho(5K) \approx 500$
and $\gamma_{\rm v}$ is the volumetric Sommerfeld coefficient. The latter
is related to the usually used molar quantity $\gamma_0$
in the present case with two KFe$_2$As$_2$ units per unit cell
by the relation 
$$\gamma_v=2\gamma_0/N_{\rm A}V_u$$ where $N_{\rm A}$ denotes 
the Avogadro's 
number 
and $V_u=204.2 $\AA $^3$ is the unit cell volume of KFe$_2$As$_2$.
Then following Hussey \cite{Hussey2005}, 
one has for the case of a (quasi)-2D system
with cylindrical Fermi surface sheets
\begin{equation}
\kappa_{\rm KWH}= \frac{A_{\rho}}{\gamma_v^2} =
\frac{72\pi\hbar}{e^2k_{\rm B}^2}\frac{ac^3}{k_{\rm F, el}^3} =\ \propto n^{-1.5} \ .
\label{Kadowaki-Woods}
\end{equation}
Notice the cancellation of many-body renormalizations on the r.h.s.\ 
and the different exponent -1.5 for the density $n$ as compared with
$\propto n^{-2}$ within  a similar expression proposed recently 
and given here for comparison also for the 2D-case, only:
\cite{Jacko09}:
\begin{equation}
\kappa_{\rm KWJ}=\frac{A_{\rho}}{\gamma_{\rm 0, el}^2}=
\frac{81\pi \hbar }{4 k_{\rm B}^2e^2n^2} \ \propto n^{-2} \ .
\label{Jacko}
\end{equation}
Since in the stoichiometric case of K-122
there is exactly one hole in the three bands (i.e.\ per formulae
unit)
which cross the Fermi energy 
one obtains for the corresponding electron density
$n_{\rm el}=4.94 \times 10^{28}m^{-3}$  and
$$k_{\rm F,el}=\sqrt{2\pi n c}=2.07\cdot 10^{-10}\mbox{m}^{-1}$$ 
for the 2D-effective
Fermi wave vector of electrons.
Inserting our value of $k_{\rm F,el}$ into Eq.\ (\ref{Kadowaki-Woods}),
one arrives at $\gamma_v=$~0.67 mJ/$K^2$cm$^{3}$ or
$\gamma_{0, \rm el} \approx 41.05$~mJ/K$^2\cdot$mol $ < \gamma_{n}$.
Using instead Eq.\ (\ref{Jacko}) one obtains a slightly smaller value
$\gamma_{\rm 0, el} \approx 36.95$~mJ/K$^2 \cdot $mol which however 
again is significantly smaller 
than our nominal value $\gamma_n\approx 94 $~mJ/K$^2\cdot$mol.
Thereby in both cases of Eqs.\ (\ref{Kadowaki-Woods},\ref{Jacko})
the experimental value
$A_\rho=3\times 10^{-2}\mu \Omega $cm has been used as
reported in Ref.\ \onlinecite{Hashimoto2010}.
Thus, our empirical value 
%within these two bounds compares very well with our
of about 60~mJ/K$^2 \cdot $mol 
\cite{Grinenko2012}  can be regarded as reasonable number.
In view of the idealized electronic structure in terms of
cylindric FSS adopted above, 
a more realistic and sophisticated multiband analysis is desirable.
To illustrate this point
we consider the simple case when all four FSS would give the same
contribution to the resistivity and to the specific heat. 
Applying the Kadowaki-Woods relation
first to such a hypothetical single FSS we would arrive finally at 
73.9~mJ/K$^2\cdot$mol in case of Eq.\ (\ref{Jacko}).
Since different individual residual resistivities
$\rho_i \ i=1-4$ lower these values,
we regard these two numbers as upper and lower bounds for 
a more realistic $\gamma_{\rm 0 el}$ somewhere in between.
More theoretical microscopic  studies 
including also the determination of the 
individual 
residual resistivities
are necessary
to improve the accuracy of 
these
Kadowaki-Woods type relations for pronounced
multband systems. 
Note that our empirical value of about 60 is very close
to the mean value 0.5(36.95+73.9)= 55.45 
and 78.57 in case of Eq.\ (\ref{Kadowaki-Woods}).
A more detailed consideration will
be given elsewhere. 
Anyhow, 
considering also available data for the in-plane penetration
depth (or the condensate density) in the superconducting state at very low $T$, 
one arrives at similar estimates (see below) and we strongly believe
that the nominal
value of about 94~mJ/K$^2$/mol given above and similar 
numbers 
found in the recent literature \cite{Fukazawa2011} as well, do
significantly {\it overestimate} the contribution from the itinerant
electrons which bear the superconductivity.

The nominal value
$\gamma_n$ 
%the effective density of states (DOS)
%at the Fermi level $N(\epsilon_{F})_{eff}$ can be estimated
%following the relations \cite{Kittel2005, Tari2003}:
%\begin{equation}\label{eq1}
%\gamma_{\rm el }= \frac{\pi^{2}k^{2}_{B}}{3}[N(\epsilon_{F})_{eff}].
%\end{equation}
%{This value 
 should be compared with the calculated quantity of
$\gamma_{b}$= 10.2 to 13.0 mJ/mol K$^2$ from DFT (density
functional theory) based 
band structure calculations
\cite{Popovich2011,Rosner,Hashimoto2010} regarded as the
unrenormalized bare quantity. The renormalization happens in two steps
at different energy scales: a first one at
high-energies is governed by the Coulomb
interaction and/or Hund's rule coupling 
and a second one at low-energies governed by the interaction
of the quasi-particles with various bosonic excitations
(phonons, paramagnons, magnons etc.). The high-energy renormalization
yields for typical transition metals a mass enhancement by a factor
two to three as evidenced by a general band squeezing as observed
for instance in ARPES measurements \cite{Sato2009,Evtushinsky2011}
or in optical measurements comparing
calculated and measured unscreened plasma frequencies.
In fact, taking a typical 122 experimental in-plane
plasma frequency of 1.55~eV \cite{remarkopt} to be compared with the calculated DFT value
of 2.56~eV to 2.58~eV \cite{Rosner,Hashimoto2010} mentioned above.
Then, 
 this high-energy 
mass enhancement can be estimated by a large factor 
of $\eta \approx 2.7$ in accord with the  high-energy band 
"squeezing" factor of  about  
2 to 3 as seen by ARPES \cite{Sato2009,Evtushinsky2011}.
Thus, one is left     with an effective quasi-particle (qp) $\gamma_{\rm qp}$
quantity of about 30 to 40 \,mJ/mol K$^{2}$
to be compared with our emprical estimate of about 60 \,mJ/mol K$^{2}$:
\begin{eqnarray}\label{eq2}
\gamma_{\rm el}&=&\gamma_{\rm qp}[1+\lambda_{\rm ph}+\lambda_{\rm sf} ],\nonumber \\
\gamma_{\rm qp}&=&\eta \gamma_b \ , \nonumber \\
\eta &\approx & \Omega^2_{\rm pl, DFT}/ \Omega^2_{\rm pl, opt} \ ,
\end{eqnarray}
where $\lambda_{\rm ph}$ 
is the electron-phonon coupling constant, $\lambda_{\rm sf}$ is
the enhancement due to spin fluctuations (antiferomagnetic 
paramagnons), and $\eta > 1$ denotes the high-energy renormalization. 
In case of Fe-based
superconductors the conventional electron-phonon interaction is weak
yielding
$\lambda_{\rm ph} \leq$ 0.2 \cite{Boeri08} which is insufficient to explain 
the large $\gamma_{\rm el}$-value
obtained from specific heat measurements.
A similar value has been found also for 
KFe$_2$As$_2$: $\lambda_{\rm ph} \approx 0.17$ (details of this 
DFT based calculation
will be given elsewhere).
%An additional enhancement
%can be caused by low energy paramagnons which is hardly to
%distinguish from the ARPES experiments \cite{Evtushinsky2011}.
Taking this into account, we may finally estimate that 
$\lambda_{\rm sf}  \stackrel{\leq}{\sim} 1$
 in the case of KFe$_{2}$As$_{2}$.
 
 \subsubsection{Penetration depth and condensate density}
The conclusion about weak electron-boson coupling
is also supported using the 
experimental value of the in-plane penetration depth extrapolated
to $T=0$: 
$\lambda_{\rm ab, L}\approx 203$~nm (measured at 50~mK)
\cite{Kawano-Furukawa11}.
Following Refs.\ \onlinecite{Drechsler08,drechsler09,Drechsler10}
one has from the {\it renormalized}
plasma frequency which enters the penetration depth
rewritten in convenient units
\begin{equation}
\Omega_{\rm pl}\mbox{[eV]}\lambda_{\rm L}\mbox{[nm]}=197.3\sqrt{DNZ_{\rm m}}, \quad 
DNZ_{\rm m} >1 \ ,
\end{equation}
where $N=n_{\rm tot}/n_s$  is the reciprocal number of the 
conduction electron density involved in the superconducting 
condensate, $Z_{\rm m} \approx (1+\lambda_{\rm tot}(0))$ describes the dynamical
mass renormalization and $D=(1+\delta )(1 +f)$ with $\delta , f >1$
describes the effect of disorder and fluctuations of competing phases.
In the clean limit one has $\delta \rightarrow 0$.
For the sake of simplicity we will ignore the 
influence of fluctuations.
Using $\hbar \Omega_{pl}=1.55$~eV for the expected unscreened 
experimental plasma
frequency, (i.e.\ the experimental high-$T$ plasma energy with no or small renormalizations
due to the electron-boson couplings) one has
\begin{equation}
2.64\frac{n_s}{n_{\rm tot}}=\left( 1+\lambda_{\rm ph}+ \lambda_{\rm sf} \right) ( 1+ \delta ) \quad ,
\end{equation}
where $\delta \sim 1/3$ measures the disorder and the reciprocal gap
amplitude
related parameter close to that in
the 
clean limit (i.e.\ $\delta \ll 1$). Furthermore  
for the sake of simplicity we will assume that all electrons are involved
in the superconducting condensate, i.e.\ $N \equiv 1$
(just for illustration see also 
the special case $n_s/n\approx 0.74$ mentioned in our remark
\cite{remnormal}).
Then, one arrives at the constraint \cite{remdisorder}:
\begin{equation}
\lambda_{\rm tot} =\lambda_{ph} +
\lambda_{\rm sf}\approx 0.97, \quad \mbox{or} \ \lambda_{\rm sf} \approx 0.8 \ .
\end{equation}
in accord with a close estimate from $\gamma_{el} \sim 60$~mJ/K$^2$ mol
and $\gamma_b \approx 10.5$~mJ/K$^2$mol from DFT calculations for the bare 
density of states \cite{Rosner}.

\begin{figure}[t]
\includegraphics[width=20pc,clip]{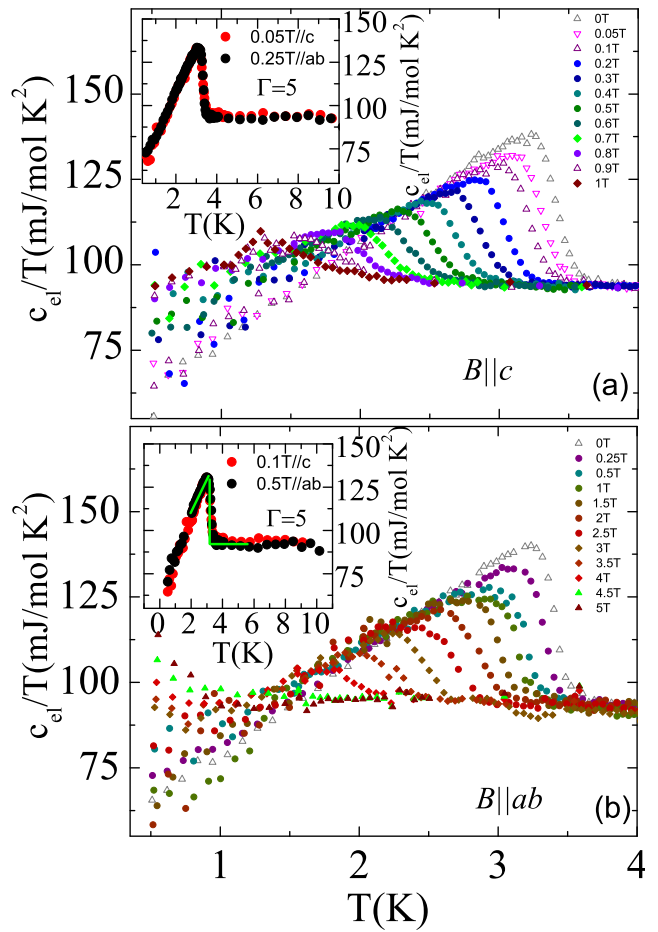}
\caption{(Color)
The electronic specific heat coefficient 
$c_{el}/T$ of KFe$_2$As$_2$ 
(after subtracting
the phonon contribution) for both directions $B \parallel c$ and $B
\parallel ab$ as shown in (a) and (b), respectively. In order to
determine the $T_{c}$ of KFe$_2$As$_2$, an entropy -
conserving construction has been used as shown with a green line in
the inset of (b). The insets of the upper and lower panel show two
data sets with the same $T_{c}$ value for the two directions
confirming our anisotropy ratio $\Gamma$ $\sim$ 5.}
\label{Fig:7}
\end{figure}
\begin{figure}[b]
\includegraphics[width=20pc,clip]{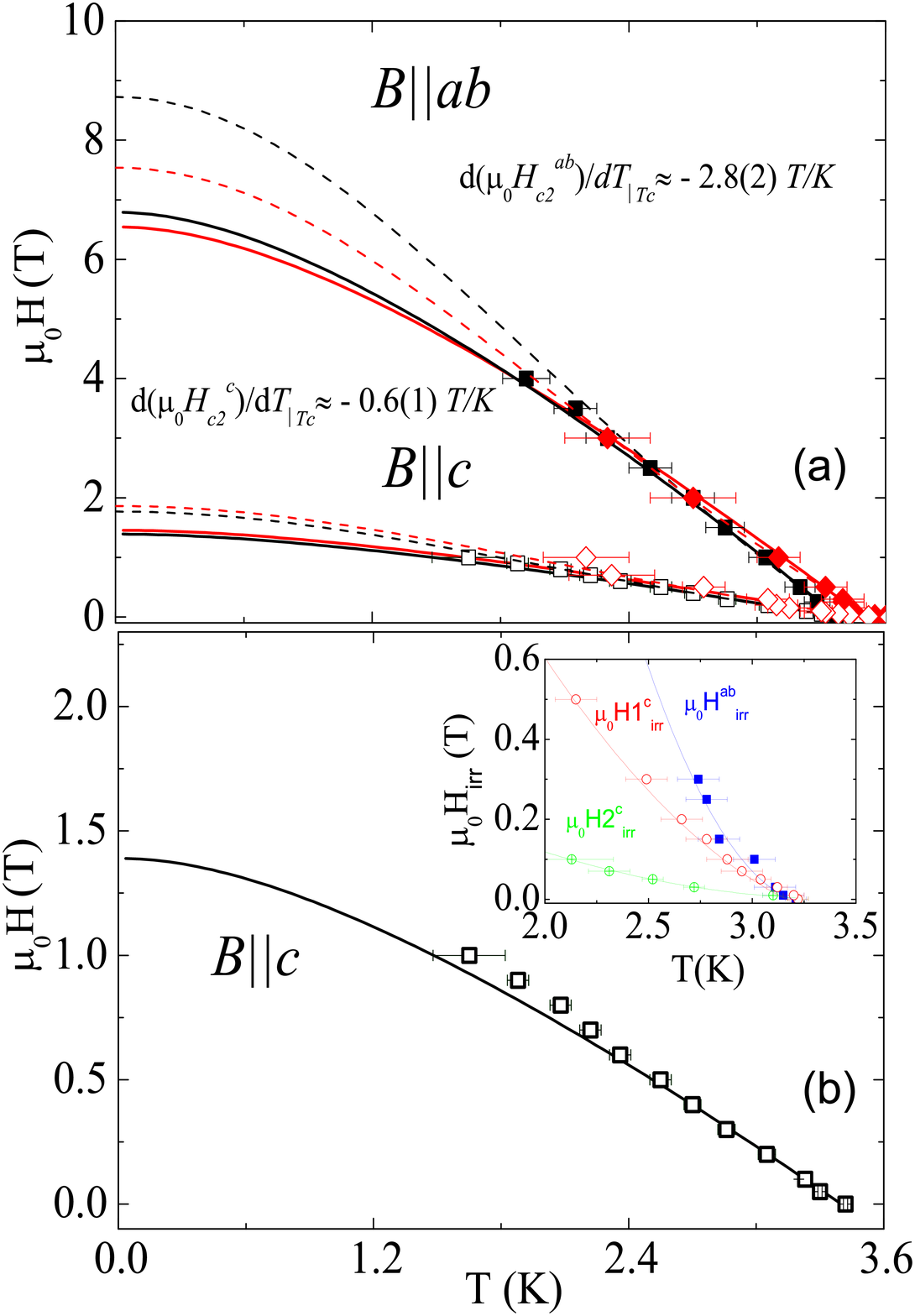}
\caption{(Color online) Phase
diagram of 
$\mu_0H_{c2}(T)$ for KFe$_{2}$As$_{2}$ with
the magnetic field applied parallel and perpendicular to the $c$
axis from specific heat (black symbols) and ac susceptibility
measurements (red symbols).
Solid lines:  theoretical curves based on the WHH model ($\alpha$ = 0 and $\lambda$$_{so}$ = 0.1).
Dashed
curves:  GL theory. Inset: $T$-dependence
of $H_{irr}$ obtained from ac susceptibility measurements
as discussed in the text.} \label{Fig:5}
\end{figure}

\subsection{The upper critical fields $H_{c2}(T)$
and their anisotropy 
%and the SC gap properties
}

Figs.\ 5-7 summarize the $T$-dependence of the specific
heat data $c_{p}$ of the investigated KFe$_{2}$As$_{2}$ single crystal for
various magnetic fields applied parallel and perpendicular to the
$ab$-plane. With increasing applied magnetic field in both
directions, the superconducting anomaly shifts and broadens
systematically to 
lower $T$ and is also reduced in
height. In an applied magnetic field of 9\,T, superconductivity is
completely suppressed for both directions of our crystal. 
In order to to analyze
the phase diagram of the field dependence of
$T_{c}$, we used an entropy-conserving construction of the
electronic specific heat to 
%accurately 
determine $T_c$ of both orientations as shown in Fig.\ 5.
Then, in a very first step
the upper critical field and its slope near $T_c$ 
can be
estimated
%evaluated 
by the Ginzburg
Landau (GL) equation~\cite{Woollam1974,Jones1964}
(strictly speaking valid near $T_c$, only):
\begin{equation}\label{eq1}
    H_{c2}=H_{c2}(0)[\frac{1-t^{2}}{1+t^{2}}],
\end{equation}
where $t = T/T_{c}$. The upper
critical field values at $T$ = 0 have been evaluated to
$\mu_0H_{c2}^{(c)}(0)$ = 1.8\,T and $\mu_0H_{c2}^{(ab)}(0)$ = 8.6\,T
and the fits are shown via dashed black lines in Fig.\ 6. In the case of 
$H_{c2}$ obtained from ac susceptibility data
(see above Fig.~\ref{Fig:1}) we derive 
at
slightly different values of
$\mu_0H_{c2}^{(c)}(0)\approx$ 1.9\,T and
$\mu_0H_{c2}^{(ab)}(0)\approx$ 7.5\,T, respectively  (dashed red curves).

It is interesting to compare the obtained and extrapolated
to $T=0$ anisotropy ratio 
for the upper critical field of 4 to 5
with that ratio for the penetration depth \cite{Ohishi11} (i.e.\ 
$\lambda_{L,ab}=194,3$~nm and $\lambda_{L,c}=$510.3~nm taken at $T=$20~mK)
which yields 2.63, only.
In fact, for a simple one-band or separable multiband
models \cite{Kogan2009} including a phenomenological
mass anistropy, one would expect
\begin{equation}
\Gamma_0 \approx \left( \frac{m_c}{m_{ab}} \right)^{1/2}=\frac{\lambda_{L,c}(0)}{\lambda_{L,ab}(0)}
=\frac{H_{c2 \parallel ab}(0)}{H_{c2 \parallel c}(0)} \ .
\end{equation}
From full relativistic 
DFT-calculations an out of plane plasma frequency of 
0.61~eV has been
obtained \cite{Rosner}
which suggests an mass anisotropy of 4.38 slightly exceeding 
the value of
3.27 for Ba-122 
\cite{Nakamura09}  
Thus, the observed anisotropy derived from the upper critical fields
exceeds this value whereas the penetration depth gives a slightly 
smaller value. 
We ascribe this small deviation
%difference 
of our empirical $\Gamma $ from the simple mass
anisotropy
to 

(i) the anisotropy of the pairing interaction and 
consequently also of the order parameter and/or oppositely 
of the depairing interaction. 
%like ($\Delta \propto \cos \Theta $) order parameter fit 
in Ref.\ \onlinecite{Ohishi11} 
which might additionally enhance
$H_{c2 \parallel ab}$  and suppress $\lambda_{L,ab}$
or vice versa the corresponding $c$-components. For instances, the anisotropic 
screening and significantly anisotropic plasma frequencies
might cause an anisotropic Coulomb pseudopotential $\mu^*$. The in-plane 
anisotropy
observed for ordered magnetic structures should in case of a magnetic 
spin fluctuation bases mechanism act in a similar way.

 (ii) Furthermore, one should take into account, that strictly
speaking, the upper critical fields and the penetration depth at $T=0$ probe
various subgroups of electrons with different Fermi-velocity dependent weights:
whereas the penetration depth probes more sensitively fast electrons
$\lambda_{L,i}^{-2}\propto \Omega^2_{pl,i}$, where $\Omega_{pl,i}$ denotes the
corresponding $i^{\rm th}$ subgroup 
plasma frequency 
and the total penetration depth is given by 
$\lambda_L^{-2}=\sum_i\lambda^{-2}_{L,i}$. In 
contrast, the upper critical fields are more
sensitive to slow electrons since $H_{c2 \parallel ab} 
\propto \left( \Phi_0/v_xv_z \right)$, where $\Phi_0$ denotes the flux quantum.  
Finally, 

(iii) anisotropic impurity scattering 
rates
might also affect $\Gamma_0$.

Another possibility to estimate roughly 
%extract 
the upper critical field $H_{c2}(0)$
is to consider the single-band Werthamer-Helfand-Hohenberg (WHH)
formula~\cite{Werthamer1966} with the Maki parameter $\alpha$ = 0 and
$\lambda_{so}$ = 0.1 which is a reasonable value of spin orbital scattering
for Fe-based superconductors \cite{Grinenko2011}. As shown with solid lines in
Fig.~\ref{Fig:5}, the specific heat and ac magnetization $H_{c2}$ data for
$B \parallel ab$ are perfectly described by the WHH model with an average slope
$-d(\mu_0H_{c2}^{(ab)})/dT \approx$ 2.8(2)\,T/K while for $H \parallel c$
the single-band WHH model   {with -$d(\mu_0H_{c2}^{(c)})/dT$ =
0.55(5)\,T/K underestimates the experimental data obtained from the specific
heat measurements (see lower panel of Fig.~\ref{Fig:5}). This deviation is
particularly visible in the case of $H_{c2}$ obtained from the ac magnetization
data where we
used -$d(\mu_0H_{c2}^{(c)})/dT$ =0.6(1)\,T/K. From these values the upper
critical fields $\mu_0H_{c2}(0)$ are found to be $\sim$ 1.4 T and $\sim$ 7 T for
 the $c$ and $ab$ direction, respectively. The observed small difference between
$H_{c2}$ obtained from the specific heat and the ac magnetization data is
expected since these methods naturally implies different criteria for $T_c$
determination.

In general, in case of multi-band superconductivity the low-$T$
$H_{c2}$-curve may exceed the single-band WHH predictions \cite{Gurevich2003}.
Therefore, we suppose that the observed deviation from the single
band WHH model is related to multi-band effects. Additionally,
indication for a two-band like behavior of our single crystal was observed in
zero field specific heat measurements (see below). Anyhow, using typical
renormalized Fermi velocities $v_F \sim 4\times 10^4$m/s
derived from preliminary ARPES-data \cite{Evtushinsky2011}
and $T_c=$ 3.5~K, one estimates also, in principle,
within a two-band approach adopting $s$-symmetry
\cite{Gurevich11,Tarantini11}, a slope-value 
\begin{equation}
H_{\rm c2, c} ' = -\frac{24 \pi k_{\rm B}^2 T_c \Phi_0 }{ 7 \zeta (3) \hbar^2 \left( c_1 v_1^2 + c_2 v_2^2 \right)} \quad ,
\label{gurevich}
\end{equation}
 where $c_1 \rightarrow c_2 \rightarrow 1/2$ and $v_F \sim \sqrt(2)v_1, \sqrt(2)v_2$
 in the case of a dominant interband pairing
 and $\zeta(3)\approx 1.202$, 
 resulting in
 -d$H_{c2}^{c}/ \mbox{d}T$=0.69~T/K near $T_c$ which is already
very
close to
our experimentally determined value
and is also in 
accord with the renormalized Fermi velocity of 4.$\cdot 10^6$cm/s
using the total bare velocity 1.77$\cdot 10^7$cm/s from the 
full relativistic (not spin polarized)
LDA calculations
and the FSS averaged
renormalizations contained in the intrinsic $\gamma_{\rm el}$-value of about 60~mJ/K$^2$mol estimated above.
In comparison, the reported} values determined
via detailed resistivity studies on KFe$_{2}$As$_{2}$ single
crystals yield lower values, i.e., $H_{c2}^{c}$ = 1.25\,T and
$H_{c2}^{ab}$ = 4.47\,T, where a low value of
$T_c$ = 2.8\,K, has been reported \cite{TERASHIMA2009}.
The anisotropy of the slopes near $T_c$ {\it as measured}
of about 5.35 is very close to the value found here: 5.09. The reported larger
value of 6.8 seems to be a consequence of the extremely high anisotropic
spin-orbit coupling $\lambda_{\mbox{\tiny s0}} =0.36$ for $B \parallel ab $ 
and $\infty $ for $B \parallel c$ adopted in Ref.\ \onlinecite{TERASHIMA2009}
in analyzing their data\cite{remterashima}. The reported 
{\it larger } absolute slope values might be interpreted 
as a hint for an impurity driven transition to an $s$-wave superconductor 
with $\langle \Delta \rangle_{\rm FS} \neq 0$
with pair-breaking (see Eq.\ (A3) in Ref.\
\onlinecite{Kogan2009}). 
From our
studies, further information about the anisotropy of
KFe$_{2}$As$_{2}$ single crystals can be obtained, which is 
$\Gamma = H^{ab} _{c2}/H^{2} _{c2} \sim $~5 (see also the 
insets of Fig.\ 5). 
Surprisingly, this anisotropy value is comparable  
with $\Gamma$-values of e.g.\ NdFeAsO$_{0.82}$F$_{0.18}$ 
\cite{Jia2008} and LaFePO \cite{JJ2008}
showing a more anisotropic electronic structure (LaFeAsO: 9.2 to 10.8
and LaFePO: 4.16 to 5.04 and might be therefore ascribed to opposite
anisotropies of the order parameter.
On the other hand, it is considerably larger than a typical
value of $\Gamma \sim$ 2 and 2.6 found for nearly optimally hole doped BaKFe$_{2}$As$_{2}$ \cite{U2009,Yuan2009},
but lower than the ones determined for SmFeAsO$_{0.85}$F$_{0.15}$
and La(O,F)FeAs thin films~\cite{U2011,Backen2008}.

The $T$-dependence of the irreversibility field $H_{irr}$ obtained from
$\chi{''}_{v}$ are shown in the inset of Fig.\ 6 (see above).
The low value of $H_{irr}^{ab}$ for $B \parallel ab$ is related with a large
anisotropy and a weak pining as expected in the case of clean single crystals.
We attribute the $H1_{irr}^{c}$ with a peak effect in the $T$-dependence of the critical current $J_c$ for $H \parallel c$
in accord with
similar observations on YBCO single crystals \cite{Giapintzakis1994}
which exhibit a rather similar anisotropy of the upper
critical field and therefore also a similar pinning behavior can be expected.

\subsection{Aspects of the electronic 
specific heat
in the superconducting state: 
the residual linear specific heat and the jump at $T_c$}

 The jump height of $\Delta c_{\rm el}/T_{c}$ 
$\approx$ 45.6~mJ/mol K$^2$ at $T_c$ is found from
%for 
our zero-field
electronic specific heat data. This value exceeds
%is higher than 
the value
which has been
%is 
reported for
%on 
a polycrystalline KFe$_{2}$As$_{2}$ sample
\cite{Fukazawa2009} but by a factor of
2
lower than the one obtained 
for the nearly optimally-hole
doped
Ba$_{0.6}$K$_{0.4}$Fe$_{2}$As$_{2}$~\cite{U2009}. 
For our estimated $\gamma_{\rm el} \sim 60$~mJ/mol$\cdot$K$^2$ 
the ratio $\Delta
c_{\rm el} / \gamma_{\rm el} T_{c}$ found to be
enhanced as compared with the use of the nominal value: near 
about 0.76 vs.\ 0.49 (see Fig.\ 4), but it
is still significantly lower
%0.37, which is much lower
than the result of the
BCS weak coupling approximation: $\Delta c_{\rm el} / \gamma_{\rm el}
T_{c}$=1.43~\cite{Bardeen1957}, which points towards a multiple band (gap)
with $s, p$- or $d$-wave nature of superconductivity. 
In particular, it is close to
the value reported for the $p$-wave superconductor SrRuO$_4$ which exhibits
0.73 \cite{Deguchi2004}.
\begin{figure}[b]
\includegraphics[width=20pc,clip]{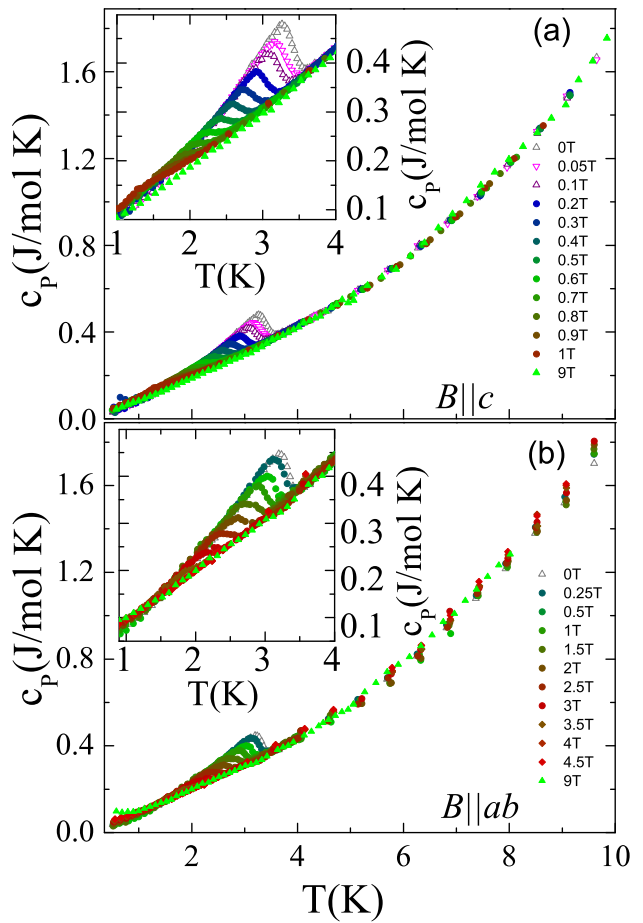}
\caption{(Color online) The $T$-dependence of the specific heat of KFe$_{2}$As$_{2}$ in various
applied magnetic fields up to 9\,T parallel to the $c$ axis (a) and
parallel to the $ab$ plane (b). The insets of the upper and lower
panel show a zoom into the superconducting state for both
directions.} \label{Fig:3}
\end{figure}

In a clean situation with negligible 
pair breaking effects, 
the reduced jump of the 
specific heat $\Delta c_{\rm el}/T_c\gamma_{\rm el}$ as compared with that
of a single $s$-wave superconductor 
might be related to  unconventional superconductivity
with nodes as discussed above and/or to pronounced multiband character
with rather different partial densities of states and gaps. Furthermore,
in relative dirty systems the unconventional superconductivity
might be driven into an $s$-wave state. To illustrate the multi-band
character we adopt here for the sake of simplicity a simple effective
weak coupling $s$-wave model like in Ref.\ \onlinecite{Pramanik2011}. Another interesting
issue we would like to address is what happened with the "extrinsic" 
linear specific heat at very low $T$. Thus, fitting the electronic
part of the specific heat within a two-band model (see the blue
curves in Fig.\ 8)
admitting also a "residual" linear Sommerfeld part, we arrive at a relative
large value of
$\gamma_{\rm res}(T\rightarrow 0)\approx 15$~mJ/K$^2$ mol
which might be related to (i)
an "extrinsic" pair breaking contribution 
somewhat 
suppressed deep in the superconducting state \cite{remnormal}.   
 Anyhow, we admit that
the adopted $s$-wave analysis might provide only an upper limit, 
since for an unconventional pairing symmetry the spectral weight at low-$T$
is enhanced. Furthermore the final density of states introduced by 
pair-breaking induced
subgap
states might also contribute to such a value. Thus, more sophisticated
multiband models including interacting pair-breaking 
impurity states are necessary
to settle this highly interesting problem. Due to its complexity it is 
however far beyond the present paper. Specific heat measurement 
below 0.2~K might be helpful to refine the value of $\gamma_{res}$.
In this contect the observation of substantial 
residual terms in
other pnictide or chalcogenide superconductors is noteworthy.
For instance in the systems FeTe$_{0.57}$Se$_{0.45}$ 
and Co-doped Ba-122
also a relatively
large (8\% and 25\% , respectively) residual linear contribution have 
been observed 
\cite{Naoyuki10,Hardy10}.

Finally, for completeness
 we list the gap values 
 obtained in the present simple model
 for analyzing 
%Furthermore, the superconducting gap properties have been
%investigated using 
the $T$-dependence of our zero-field
specific heat measurements down to 400\,mK. The normalized
zero-field electronic specific heat $c_{\rm el}/\gamma_{\rm n}T$ is shown in
Fig.\ 8. First, we compare our data to the single-gap BCS theory
(i.e.\ the weak coupling approach using $\Delta_0$/$k_{B}T_{c}$ =
1.76 at $T_c$) and find that a single BCS-gap cannot be reconciled with
our experimental data. From the plot in Fig.\ 8 it is evident that the
specific heat jump is much lower than the BCS weak-coupling value
for an $s$-wave superconductor. Below T$_{c}$, systematic deviations
of the theoretical curve from the experimental data have been
observed at both low-$T$ and around $T_c$.
This clearly 
indicates that $c_{\rm el}/\gamma_{\rm el}T$ of KFe$_2$As$_2$
cannot be described by the weak coupling BCS (single-)gap. 
%which is in agreement with 

Since a single-gap scenario cannot describe our data, we applied a
phenomenological two-gap model in line with multi-gap
superconductivity in many compounds of the
FeAs family reported by
various experimental  and
theoretical studies
%approaches 
on different compounds within this family
\cite{Mazin2008,Yashima2010,Fukazawa2009,Shermadini10}. 
We have analyzed our
data utilizing the generalized $\alpha$-model which has been
proposed to account for the thermodynamic properties in multi-band,
multi-gap superconductors like e.g. MgB$_2$ \cite{Bouquet2001}.
We remind the reader that in this approach
the one-band expression:
\begin{equation}\label{entropy}
    \frac{S}{\gamma_{\rm el}T_{c}}=-\frac{6\Delta_{0}}{\pi^{2}k_{B}T_{c}}\int_{0}^{\infty}[f\ln f+ (1-f)\ln (1-f)]dy,
\end{equation}
\begin{equation}\label{SH}
     \frac{c_{\rm el}}{\gamma_{\rm el}T_{c}}= t\frac{d(
     \frac{c_{\rm el}}{\gamma_{\rm el}T_{c}})}{dt} \  ,
\end{equation}
is straightforwardly generalized to the two-band case and entropy conservation
is adopted for each band. In Eq.\ (\ref{entropy}) the Fermi-function is 
denoted by
$f$ = $[$exp($\beta E$ + 1$)]$$^{-1}$, $\beta$ =
($k_{B}T$)$^{-1}$ and the energy of the quasiparticles is given by
$E$ = $[\epsilon^{2} + \Delta^{2}(t)]^{0.5}$ with $\epsilon$ being
the energy of the normal electrons relative to the Fermi surface.
The integration variable is y = $\epsilon$/$\Delta_0$. (S) and (C) is the 
thermodynamic properties and $t=T/T_{c}$ is the reduced temperature. 
In Eq.\  (\ref{entropy} the
scaled gap $\alpha = \Delta_0/k_BT$ is the only adjustable fitting
parameter. The temperature dependence of the gap is determined by
$\Delta(t) = \Delta_0\delta(t)$, where $\delta(t)$ is 
approximately described by the data taken
%obtained 
from the table in Ref.~\cite{muehlschlegel59}. In case of two gaps the
thermodynamic properties are obtained as the sum of the
contributions from the two gaps, i.e., $\alpha_1$ =
$\Delta_1(0)/k_BT_c$ and $\alpha_2$ = $\Delta_2(0)/k_BT_c$ with
their respective weights $\gamma_1/\gamma_{\rm m}$ and
$\gamma_2/\gamma_{\rm el}$.

\begin{figure}
\includegraphics[width=20pc,clip]{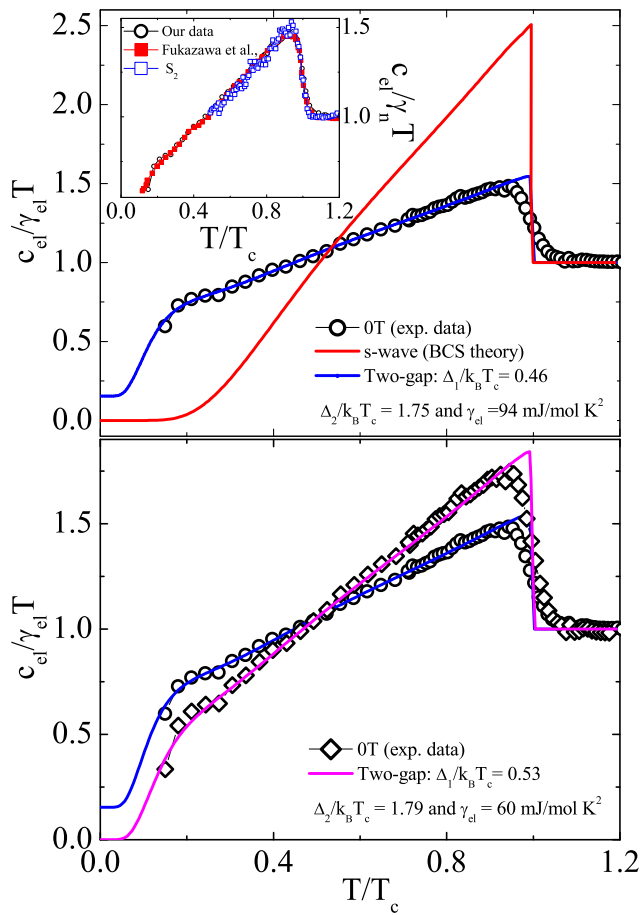}
\caption{(Color online) Upper panel: Fit under the assumption
of no extrinsic contribution to the linear specific heat (see also 
Figs.\ 3 and 4).
The
normalized superconducting electronic specific heat
$c_{el}/(\gamma_{\rm el}T)$ of KFe$_{2}$As$_{2}$ as a function of the
reduced temperature $t = T/T_{c}$. The red line represents
the theoretical curve 
for the
single-band weak coupling BCS case
($\Delta_0/k_{B}T_{c}$ = 1.76). The blue line shows the curve of the
nodeless weakly coupled
two-gap model fit; for details see text. The inset shows our
electronic specific heat data in comparison with data by Fukazawa 
{\it et al.}, \cite{Fukazawa2011} together with an another sample 
which
has been shown 
that our $T_{c}$ is similar to the
investigated sample. Lower panel: fit 
%under the assumption 
assuming
that
a significant contribution to the linear specific heat is not intrinsic:
using e.g.\ $\gamma_{\rm el}=60$~mJ /mol K$^2$ taken 
from Ref.\ \onlinecite{Grinenko2012}
(see Figs.\ 3 and 4.).} \label{Fig:6}
\end{figure}

To calculate the theoretical curves c$_{\rm el}/\gamma_{\rm el}T$ the
parameters $\Delta_1$, $\Delta_2$, their respective ratios
$\gamma_1$ and $\gamma_2$ and the ratio $\gamma_{res}/\gamma_{\rm el}$
are left for free in fitting ($\gamma_{res}$ represents the 
non-negligible
residual value at
low-$T$). The best description of the experimental data is
obtained using values of $\Delta$$_{1}$/$k$$_{B}$T$_{c}$ = 0.46 and
$\Delta$$_{2}$/$k$$_{B}$T$_{c}$ = 1.75. The calculated specific heat
data are represented by the solid blue line in Fig.\ 8 (upper panel. 
Anyhow, small relative jumps are not compatible
with the strong coupling scenario estimated in Fig.\ 4 for the case of no 
extrinsic contributions. Therefore we performed a second analysis
where the effective extrinsic linear contibution necessary for a weak coupling
scenario has been subtracted fromn the row data. The result is shown in 
the lower panel of Fig.\ 8. Then both gaps do slightly
increase: 1.8~K and 6.2~K.

Naturally, the obtained  
 gap values  are smaller than
the largest gap in e.g.\ the optimally hole doped
Ba$_{0.6}$K$_{0.4}$Fe$_{2}$As$_{2}$ as observed by ARPES
investigations~\cite{Ding2008}
but comparable to 
the two-band $s$-wave fit for the penetration depth data 
(H $\parallel c$: 1.28~K and 5.57~K but 
do not clearly exceed
 the corresponding values
for the isomorphic compound RbFe$_2$As$_2$ 
system
with a slightly lower 
$T_c$-value of 2.52~K, only: 1.74~K  
and 5.7~K. \cite{Shermadini10}. In our opinion this might
reflect the presence of nodes in the superconducting 
order parameter of KFe$_2$As$_2$. Anyhow, a detailed
comparison of these two closely related systems would be very interesting,
especially, if in fact it would be confirmed that 
the symmetry of the order parameter would be
different.

Although a clear picture is still missing for the case
of KFe$_2$As$_2$, it is important to emphasize that our system
definitely underlies multiband superconductivity, 
probably in the weak coupling regime.
However, from specific heat data alone it is difficult to be sure
whether nodes exist or not, since in the case of multi-band
superconductivity low-energy quasi-particle excitations can be
always explained by the contribution from 
an electron group with a
 small gap. We believe
that further experimental studies 
%using different methods 
such as
specific heat well
below 400\,mK,  ARPES and transport investigations at very low
$T$, will be helpful to elucidate 
%finally resolve 
the nature
of superconductivity in KFe$_2$As$_2$.

\section{Conclusions}

In summary, KFe$_{2}$As$_{2}$ 
%with a superconducting transition
%temperature $T_{c}\approx$ 3.5\,K 
was investigated by ac
susceptibility and low-$T$ specific heat measurements
on high-quality
single crystals grown by 
%using 
a self-flux technique. The 
specific heat jump was found to be $\Delta c_{\rm el}/T_c$ $\sim$ 45.9
mJ/mol K$^2$ and the nominal Sommerfeld coefficient $\gamma_{\rm n}$
= 94(3) mJ/mol K$^2$. 
However, several theoretical considerations 
including two recently proposed modified Kadowaki-Woods relations
as well as the 
observation of a significant linear in $T$
residual term point to a significantly 
{\it smaller} value for the itinerant
quasi-particles of about 60 mJ/K$^2$ mol.
This way the strongly correlated 
"heavy-fermion-like"
scenario suggested for K-122 in the literature
should be revisited. In this context 
 the elucidation of the
"external" 
subsystem responsible for that difference is a challenging problem
to be considered elsewhere. The total electron-boson coupling constant
$\lambda_{\rm tot}=\lambda_{\rm ph}+\lambda_{\rm sf} \sim 1$ avaraged over all Fermi surfaces
excludes strong coupling. The calculated weak electron-phonon coupling 
of about 0.17
points to a 
dominant spin-fluctuation mechanism and unconventional superconductivity.

The magnetic phase diagram has been
studied yielding values for the upper critical fields
$\mu_0H_{c2}^c(0)\approx$ 1.4\,T and $\mu_0H_{c2}^{ab}(0)\approx$ 7\,T for the
$c$ axis and $ab$ plane, respectively.
%, extrapolated using a single-band WHH model. 
The resulting anisotropy of
KFe$_2$As$_2$ near $T_c$
lies around $\Gamma= H_{c2}^{(ab)}/H_{c2}^{(c)}\sim$
5 which slightly exceeds the mass anisotropy as derived
from DFT-electronic structure calculations as well as the anisotropy of 
the penetration depth.
 Additionally,
the $T$-dependence of our zero-field
electronic specific heat $c_{el}$ cannot be described within
single-band weak-coupling BCS theory. 

For a full understanding of the gap structure of KFe$_2$As$_2$
as well as of the high values of $c_{\rm el}/\gamma_{\rm el} T$ at low
temperatures, further specific heat measurements
at very low $T < 400$~mK and/or low-$T$ ARPES and transport studies
will be helpful. Finally, the irreversibility field $H_{irr}$ derived 
from ac susceptibility data has been investigated. The double-maximum in
$\chi{''}_{v}(T)$  for $H||c$ suggests the presence of a peak effect in the
$T$-dependence of the critical current.

\vspace{0.5cm}

\begin{acknowledgments}

The authors thank V. Zabolotnyy, 
%L.\ Boeri, 
A.\ Chubukov,
G.\ Fuchs and S. Borisenko
for fruitful discussions and M. Deutschmann, S. M\"{u}ller-Litvanyi,
R. M\"{u}ller, J.\ Werner, S.\ Pichl, and S.\ Gass and K. Nenkov for
technical support. This project was supported by the DFG through SPP
1458 and Grants No. GR3330/2 and BE1749/13. SW acknowledges support by DFG under the Emmy-Noether program (Grant No. WU595/3-1).
Financial support by the
Pakt for Forschung at the IFW-Dresden is also acknowledged by V.G and S.-L.D.
S.J.\ thanks the Foundation for Fundamental Research on
Matter (The Netherlands) for financial support.

.
\end{acknowledgments}


\begin{thebibliography}{100}
\bibitem{remdisorder}
 We stress that the small disorder adopted in the present analysis using 
 Eqs.\ (5,6) is  not applicable
to our samples but only to that superclean sample of Ref.\ \onlinecite{Kawano-Furukawa11}
for which the small penetration depth has been reported.
\end{thebibliography}
\end{document}